\documentclass[aps,prb,nobibnotes,twocolumn,superscriptaddress,bibliography]{revtex4-2}

\usepackage{amsfonts}
\usepackage{mathrsfs}
\usepackage{amsmath}% needed for subequations
\usepackage{color}
\usepackage{graphicx}
\usepackage{bm}% bold maths
\usepackage{amssymb}
\usepackage{xspace}
\usepackage{epstopdf}
\usepackage{dcolumn}% Align table columns on decimal point
\usepackage{longtable}
\usepackage{multirow}
\usepackage{float}
\usepackage{comment}

\providecommand{\tabularnewline}{\\}

\usepackage[colorlinks=true, letterpaper=true, pdfstartview=FitV, linkcolor=blue, citecolor=blue, urlcolor=blue]{hyperref}

\makeatother

\begin{document}

\title{Crossed real nodal-line phonons in gold monobromide}

\author{Yilin Han}
\affiliation{Key Lab of advanced optoelectronic quantum architecture and measurement (MOE), Beijing Key Lab of Nanophotonics $\&$ Ultrafine Optoelectronic Systems, and School of Physics, Beijing Institute of Technology, Beijing 100081, China}

\author{Yichen Liu}
\affiliation{Key Lab of advanced optoelectronic quantum architecture and measurement (MOE), Beijing Key Lab of Nanophotonics $\&$ Ultrafine Optoelectronic Systems, and School of Physics, Beijing Institute of Technology, Beijing 100081, China}

\author{Chaoxi Cui}
\affiliation{Key Lab of advanced optoelectronic quantum architecture and measurement (MOE), Beijing Key Lab of Nanophotonics $\&$ Ultrafine Optoelectronic Systems, and School of Physics, Beijing Institute of Technology, Beijing 100081, China}

\author{Cheng-Cheng Liu}
\email{ccliu@bit.edu.cn}
\affiliation{Key Lab of advanced optoelectronic quantum architecture and measurement (MOE), Beijing Key Lab of Nanophotonics $\&$ Ultrafine Optoelectronic Systems, and School of Physics, Beijing Institute of Technology, Beijing 100081, China}

\author{Zhi-Ming Yu}
\email{zhiming\_yu@bit.edu.cn}
\affiliation{Key Lab of advanced optoelectronic quantum architecture and measurement (MOE), Beijing Key Lab of Nanophotonics $\&$ Ultrafine Optoelectronic Systems, and School of Physics, Beijing Institute of Technology, Beijing 100081, China}

\begin{abstract}
Spacetime inversion symmetry can generate intriguing types of spinless excitations in crystalline materials. Here, we propose a topological phase protected by spacetime inversion symmetry--the crossed real nodal line (RNL) in the phonon spectrum of gold monobromide (AuBr). In AuBr, there exist four straight nodal lines, which are linked by a crossed nodal line formed by two lower bands. Remarkably, each adjacent two of the four straight nodal lines is a pair, forming a crossed RNL with nontrivial real Chern number. Such configuration and pairing mode of RNL have never been reported. The crossed RNL exhibits unique surface and hinge states distinguished from that of the conventional RNLs. The symmetry protection and the transformation under the symmetry-preserving strain of the crossed RNL are also investigated. Our results open the door to a new class of topological states, and predict its realization in experimentally synthesized material.
\end{abstract}
\maketitle

%{\emph{\textcolor{blue}{Introduction.--}}}
\section{\label{sec:level1}Introduction}
The crystalline materials can exhibit various excitations around band degeneracies in their energy spectrums, such as electronic and phonon band structure \cite{armitage2018weyl,RN1178,lv2021experimental,bouhon2020nonabelian,yu2022encyclopedia,zhang2022encyclopedia,zhang2023encyclopedia,
zhang2018doubleweyl,xia2019symmetryprotected,zhu2022phononica,PhysRevLett.124.105303}.
%wehling2014dirac, yan2017topological,yu2017topologicala,bernevig2018recent,yang2018symmetry,wu2018nodalb,gao2019topological,
%lv2021experimental, zhang2018doubleweyl,xia2019symmetryprotected,peng2022multigap,yu2022encyclopedia,zhang2022encyclopedia,zhang2023encyclopedia}.
The most common degeneracy is the nodal point \cite{wan2011topological,PhysRevLett.107.186806,fang2012multiweyla,young2012diraca,wang2012diraca,wang2013threedimensionala,RN1300,yang2014diraca,
PhysRevB.93.241202,weng2015weyl,PhysRevX.6.031003,PhysRevB.101.205134}, which has zero-dimensional (0D) manifold, and  widely appears in 1D, 2D and 3D systems.
With additional symmetries, the degeneracy with 1D manifold can appear, leading to the topological nodal-line phases \cite{weng2015topological,fang2015topologicala,kim2015diraca,yu2015topological,bian2016topological,PhysRevLett.118.016401,xu2017topological,RN2332,PhysRevB.99.121106}.

Generally, the nodal line occurs in a mirror plane formed by two bands with opposite mirror eigenvalues \cite{weng2015topological,xu2017topological}.
Since the mirror can quantize the Berry phase into $0$ and $\pi$, the mirror nodal line hosts a $Z_2$ topological charge, and is  topologically protected by $\pi$ Berry phase \cite{xiao2010berryb,RevModPhys.88.035005}.
In addition, the nodal line can widely exist in the spinless systems with spacetime inversion symmetry (${\cal{PT}}$), where the  Berry phase also is quantized \cite{schnyder2008classification,zhao2016unified}.
Interestingly, the ${\cal{PT}}$ symmetry can protect another $Z_2$ topological quantity, known as second Stiefel-Whitney number $w_2$ or the real Chern number $\nu_R$ \cite{zhao2017pta,ahn2018banda,bouhon2019wilson,fang2016topologicalb,ahn2018banda,ahn2019stiefel,
wang2020boundarya,chen2022secondorder}.
Therefore, the nodal line in ${\cal{PT}}$-symmetric spinless systems can be classified as the conventional one that is solely protected by $\pi$ Berry phase, and the real nodal line  (RNL) that is protected by both $\pi$ Berry phase and the real Chern number $\nu_R$.

\begin{figure}[t]
\includegraphics[width=0.48\textwidth]{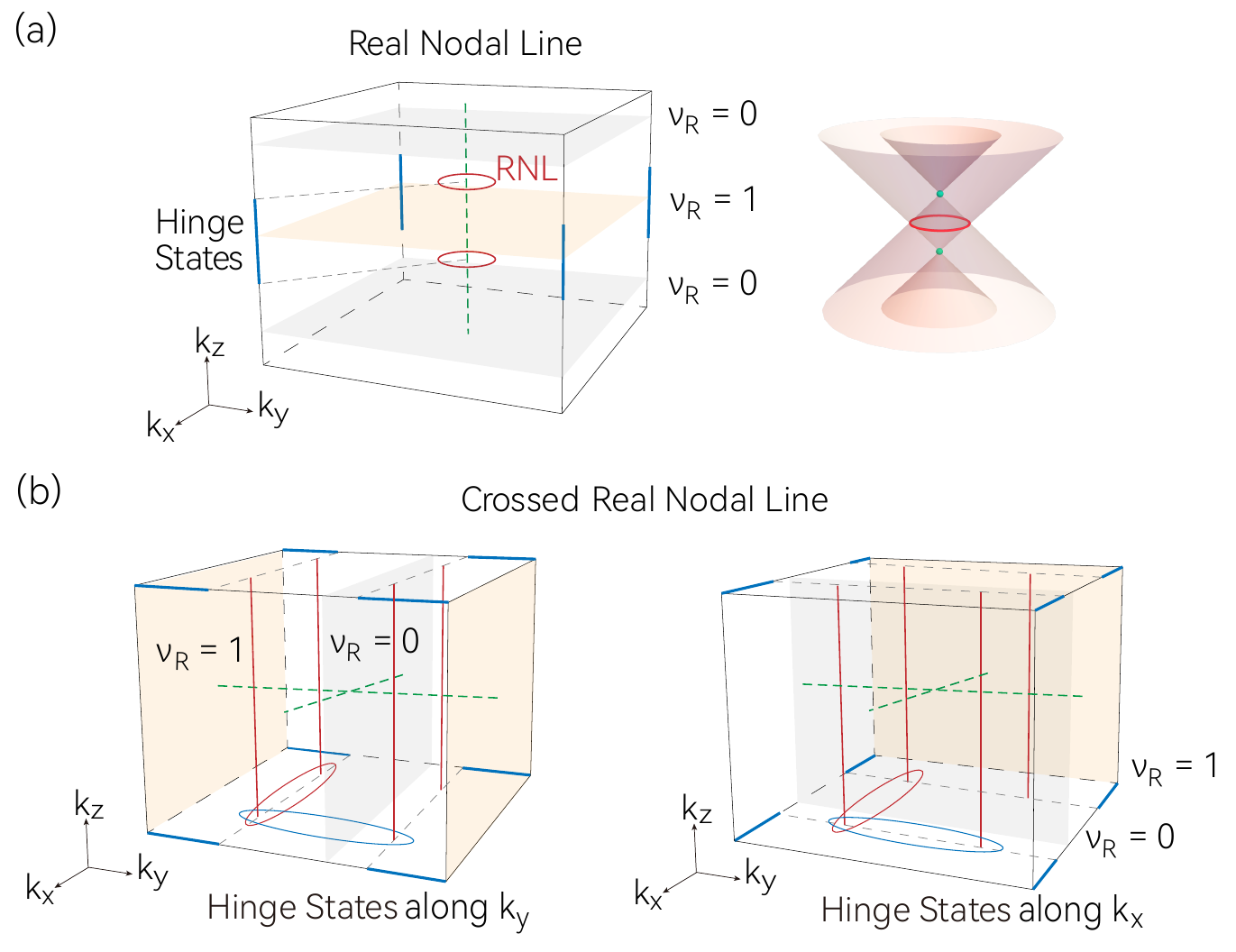}% Here is how to import EPS art
\caption{\label{RNL} Schematic figures showing two types of RNL configuration. (a) Conventional RNL configuration, where the pair of real nodal loops formed by two middle bands (red circle) are linked by another nodal line formed by two bottom bands (green dotted lines), with its band structures shown in the right.
(b) Crossed RNL  proposed in this work. All the four straight nodal lines (red lines) are conventional nodal line featuring nontrivial $\pi$ Berry phase.
However, they are linked by a  crossed nodal line (green dotted line), and each adjacent two of the four straight nodal lines \emph{together}  (wrapped  by  blue and red circles) form a RNL with trivial Berry phase but nontrivial real Chern number $\nu_R$.
Thus,  the systems with  crossed RNL can exhibit  hinge states in both $k_x$ and $k_y$ directions. In contrast, the conventional RNL state has  hinge states in only one direction.}
\end{figure}

The RNL, distinguished from the conventional nodal line, is not alone but must be linked by an additional nodal line formed by lower bands \cite{fang2016topologicalb,ahn2018banda}.
The linked structure is a typical characteristic of the RNL.
Besides, the RNL must occur in pairs, due to the nontrivial real Chern number \cite{ahn2019stiefel}.
The argument is similar to that of a Weyl point with finite Chern number \cite{wan2011topological,yu2019circumventing}.
Consider a 2D slice in the Brillouin zone (BZ) of a 3D  ${\cal{PT}}$-symmetric  spinless system.
The real Chern number of this slice is changed by one when it moves across a RNL.
Since the BZ is periodic, the slice must cross another RNL to make sure that its real Chern number is not changed when it return to its original position.
Since the real Chern insulator is a second-order topological insulator \cite{benalcazar2017quantized,xie2021higherorder,
schindler2018higherorderb,
bouhon2020geometric,song2017dimensional,langbehn2017reflectionsymmetric,schindler2018higherordera,wang2018hightemperature,fang2019new,
sheng2019twodimensionala,zhang2023magnetic,gonghidden,wang20243d,xu2019higherordera,chen2021fieldtunable}
hosting a corner state, this argument also indicates the appearance of the hinge states connecting two RNLs in the topological RNL systems \cite{chen2022secondorder}.

The material candidate of RNL is firstly proposed by Anh $\textit{et al.}$ in Ref. \cite{ahn2018banda}.
Subsequently, the RNL is predicted to exist in a few  electronic systems with negligible spin-orbit coupling \cite{chen2022secondorder,lee2020twodimensionalb}, and artificial periodic systems \cite{xue2023stiefelwhitney,xiang2024demonstration,ma2024observation,zhang2022design}.
Moreover, some theoretical works have constructed suitable model to study the RNL \cite{du2022weyl,gao2023engineering,zhao2020topological,li2023layer}.
However, the RNL in all current studies has the same configuration that is illustrated in Fig. \ref{RNL}(a).
The RNL occurs when four bands successively intersect, and it is the loop formed by the two middle bands and is linked by another nodal lines formed by the two  bottom (top) bands [see Fig. \ref{RNL}(a)].
For the conventional RNL state, it generally has  hinge states in only one direction, as shown in Fig. \ref{RNL}(a).

In this work, we propose a  type of RNL that is different from the conventional RNL, and predict its realization in the phonon spectrum of experimentally synthesized 3D materials AuBr.
We show that there are four straight nodal lines passing through the entire BZ in the phonon spectrum of AuBr.
While the four straight  lines are  conventional nodal lines, each adjacent two of them is paired to form a RNL, possessing  nontrivial $\nu_R=1$.
A typical feature  of the RNL is the linking structure.
For the RNL in AuBr, we find that the four straight nodal lines indeed are linked by a crossed nodal line formed by lower bands.
Thus, we term such nodal lines as crossed RNL, and its configuration is illustrated in Fig. \ref{RNL}(b).
For crossed RNL, we have a pair of RNLs in two directions, say, $k_x$ and $k_y$  directions [see Fig. \ref{RNL}(b)].
Consequently, the crossed RNL can have hinge states in both  $k_x$ and $k_y$ directions.
This is completely different from the conventional RNL.
Moreover, the surface state of AuBr also is different from that in previously reported RNL materials.
Furthermore, we find that the crossed RNL has unique evolutionary behavior under symmetry-preserving strain, namely, it disappears when the strain is within a certain range, but \emph{reappears} by further increasing the strain.

%{\emph{\textcolor{blue}{Methods.--}}}

\begin{figure}[t]
\includegraphics[width=0.4\textwidth]{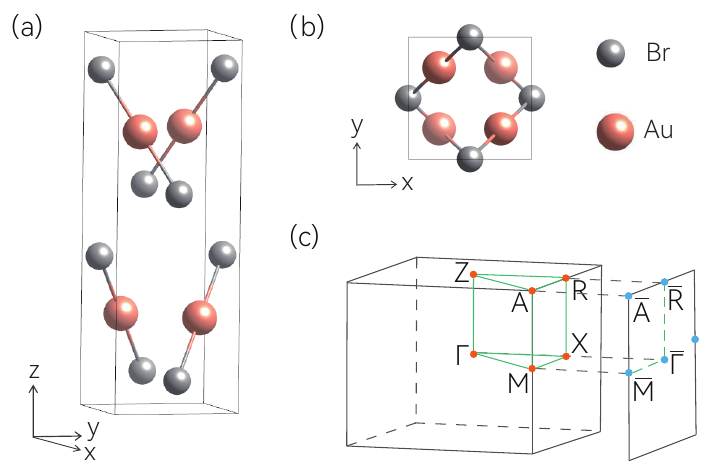}% Here is how to import EPS art
\caption{\label{crystal} (a) Side view and (b) top view of the crystal structure for 3D AuBr. (c) The bulk BZ and the (010) surface BZ for AuBr.}
\end{figure}

\section{Phonon spectrum and  crossed RNL}
The 3D AuBr has been experimentally synthesized in 1978 \cite{janssen1978phasea,janssen1978crystala}.
The crystalline structure of AuBr is plotted in Figs. \ref{crystal}(a-b), from which one observes that there are four Au and four Br atoms in a unit cell.
The AuBr has a tetragonal lattice and belongs to the space group No. 138 ($P4_2/ncm$).
From our first-principles calculation, the relaxed lattice constants for AuBr are $a = b = 4.21 $ \AA~ and $c = 12.30 $ \AA~.
The  bulk BZ and  (010) surface BZ of AuBr are illustrated in Fig. \ref{crystal}(c).
The symmetry operators belonging to AuBr and relevant here are spatial inversion (${\cal{P}}$), four-fold  rotation ($C_{4z}$),  several mirrors ($M_{z}$, $M_x$, $M_y$, $M_{110}$, and $M_{\bar{1}10}$), and time-reversal symmetry (${\cal{T}}$).

\begin{figure}[t]
\includegraphics[width=0.4\textwidth]{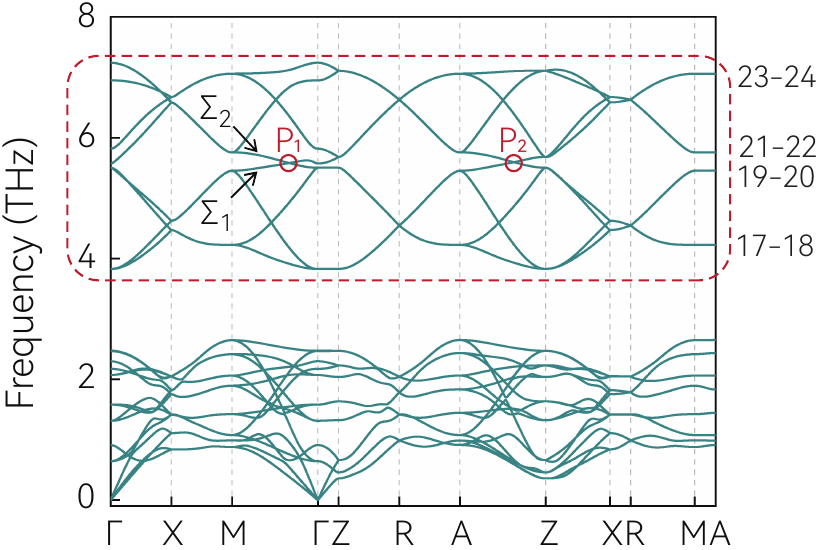}% Here is how to import EPS art
\caption{\label{phonon} Calculated phonon spectrum of 3D AuBr. The group of bands from 17$th$ to 24$th$ are our focus (circled by red). Here, two band crossing points $P_1$ and $P_2$ on paths $\Gamma$M and AZ are marked in red, which belong to the four straight nodal lines passing through the entire BZ. Phonon bands along ZR path  are doubly degenerate, forming a crossed nodal line, due to the $C_{4z}$ in 3D AuBr. }
\end{figure}

\global\long\def\arraystretch{1.2}%
\begin{table}[b]
\caption{Part of the character table of point group $C_{2v}$. Only two irreducible representations are  presented.}\label{tab:table1}
\begin{ruledtabular}
\begin{tabular}{ccccc}
$C_{2v}$ & \textit{E}  & $C_{2,110}$ & $M_z$ & $M_{110}$ \tabularnewline
\hline
$\Sigma_1$ & 1 & 1 & 1 & 1 \tabularnewline
$\Sigma_2$ & 1 & -1 & 1 & -1 \tabularnewline
\end{tabular}
\end{ruledtabular}
\end{table}

\begin{figure}[t]
\includegraphics[width=0.48\textwidth]{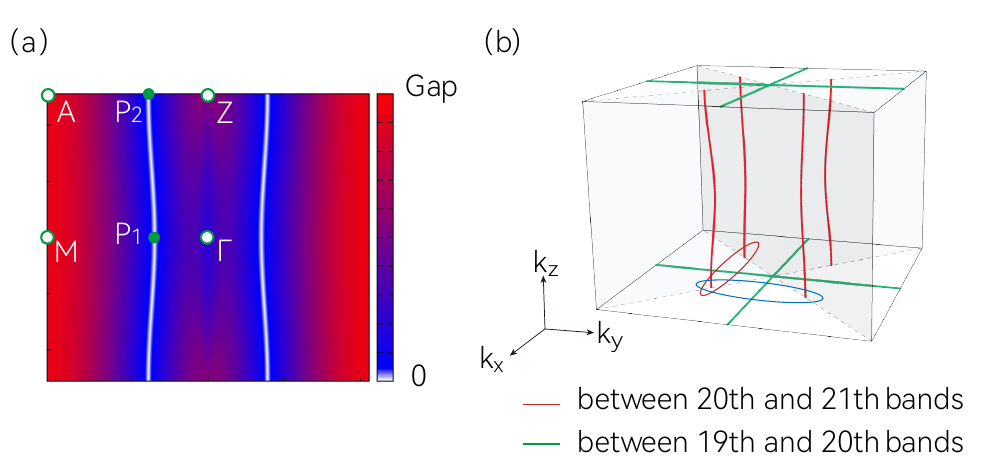}% Here is how to import EPS art
\caption{\label{CRNL of AuBr} Schematic figure showing the location of the nodal lines (a) in the $k_{110}$ plane,  and (b) in the BZ, formed by two middle bands (20$th$ and 21$th$ bands).  %The two green dots $P_1$ and $P_2$ in (a) are the crossing points on paths $\Gamma$M and AZ in phonon spectrum.
Each adjacent two of the four straight  nodal lines in (b) can be paired to form a RNL  with nontrivial $\nu_R=1$, connected by the  crossed nodal lines formed by 19$th$ and 20$th$  bands along the ZR path  (green lines).}
%(a) Schematic figure showing the location of the nodal lines in the $k_{110}$ plane of AuBr formed by two middle bands (20$th$ and 21$th$ bands), where the two green dots $P_1$ and $P_2$ are the crossing points on paths $\Gamma$M and AZ in phonon spectrum.
%(b) Schematic figure showing the location of the four straight  nodal lines  formed by 20$th$ and 21$th$  bands (red curves). Each adjacent two of the four straight  nodal lines can be paired to form a RNL with nontrivial $\nu_R=1$, connected by the  crossed nodal lines formed by 19$th$ and 20$th$  bands along the ZR path  (green lines).

\end{figure}

The phonon spectrum of AuBr along high-symmetry paths is plotted in Fig. \ref{phonon}, from which one notes that there are a group of bands (17$th$-24$th$ phonon bands) that is well separated from others in frequency.
We first investigate the two band crossing points on paths $\Gamma$M and AZ formed by the two middle bands (20$th$ and 21$th$ bands) of this group of bands, which are labeled as $P_1$ and $P_2$ [see Fig. \ref{phonon}].
Since the AuBr has both ${\cal{P}}$ and ${\cal{T}}$, $P_1$ ($P_2$) can not be isolated but a point on a nodal line \cite{weng2015topological}.
As discussed by Zhang et. al. \cite{zhang2017typeii}, the  position of the nodal line can be determined by calculating the representations of the bands forming the point.
Here,  we calculate the band representations of the 20$th$ and 21$th$ bands on path $\Gamma$M to determine the position of the nodal line including $P_1$.
For path $\Gamma$M, its litter point group is  $C_{2v}$, indicating that path $\Gamma$M has two mirrors: $M_z$ and $M_{110}$.
The band representations of the  two relevant bands are respectively obtained as $\Sigma_1$ and $\Sigma_2$, as shown in Fig. \ref{phonon}.
From the character table of  $C_{2v}$ (see Table \ref{tab:table1}), we find that $\Sigma_1$ and $\Sigma_2$ have same eigenvalues for  $M_z$ but opposite eigenvalues for  $M_{110}$.
This means that the nodal line containing  $P_1$ point may be within the $M_{110}$ plane, which also can be inferred from the effective Hamiltonian of the $P_1$ point \cite{yu2022encyclopedia}.
By a careful scanning on the $k_{110}$ plane, we find that there indeed exist a nodal line passing through the entire BZ, which includes both $P_1$ and $P_2$ points, as shown in Fig. \ref{CRNL of AuBr}(a).
Due to the $C_{4z}$ symmetry,  there are  four such nodal lines in total in the BZ [see Fig. \ref{CRNL of AuBr}(b)].

The four straight nodal lines in $k_{110}$ ($k_{\bar110}$) plane are conventional nodal lines protected by both $M_{110}$ ($M_{\bar110}$) and ${\cal{PT}}$ symmetries, and hosting nontrivial $\pi$ phase.
However, below the four nodal lines, the  phonon bands along the ZR path also are doubly degenerate [see Fig. \ref{phonon}].
The configuration of the  four  straight nodal lines and the nodal lines along ZR path are plotted in Fig. \ref{CRNL of AuBr}(b), which is reminiscent of the linking structure of the RNL if we take each adjacent two of the four straight nodal lines as a pair  [see  Fig. \ref{CRNL of AuBr}(b)].
To directly demonstrate the real topology of the paired straight nodal lines, we calculate the Wilson loops at two slices that are separated by them for the four bands from 17$th$ to 20$th$.
Here, we consider the two slices at $k_y = 0$ and $k_y =\pi$ planes.

\begin{figure}[t]
\includegraphics[width=0.47\textwidth]{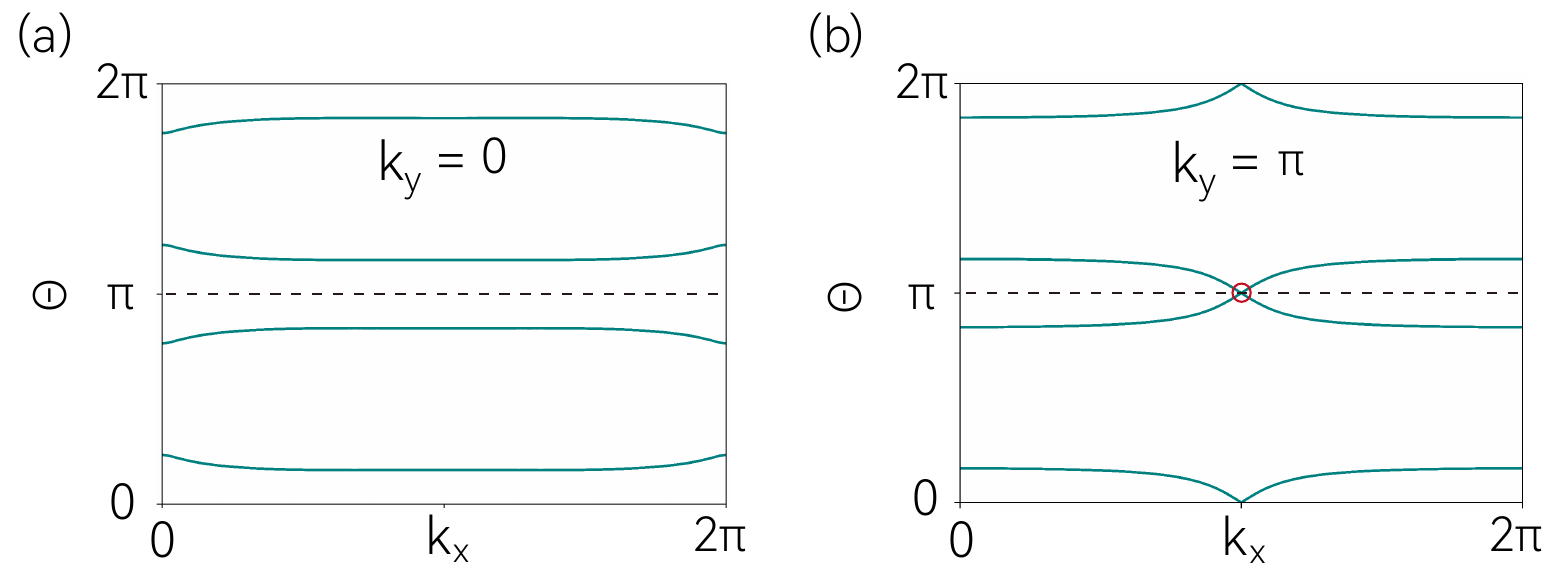}% Here is how to import EPS art
\caption{\label{wilsonloop} (a) The Wilson loop of the 17$th$ to 20$th$ bands calculated in the $k_y = 0$ plane and (b) $k_y = \pi$ plane along the $k_z$ direction. The zero (even) and one (odd) crossing point at $\theta = \pi$ indicates the planes $k_y = 0$ and $k_y = \pi$ having trivial and nontrivial real Chern number, respectively. }
\end{figure}

The calculated Wilson loops are presented in Fig. \ref{wilsonloop}, from which one notices that  for $k_y = 0$ and $k_y = \pi$ planes, there are even (zero) and odd (one) crossing points at $\pi$ axis, respectively.
This indicates that the  $k_y = 0$  ($k_y = \pi$) plane is trivial (nontrivial) with real Chern number $\nu_R = 0$ ($\nu_R = 1$).
The difference  of the $\nu_R$ of $k_y = 0$ and  $k_y = \pi$ planes directly shows the two straight nodal lines at $0<k_y<\pi$ region  as a whole is a RNL with $\nu_R = 1$.
Again, the $C_{4z}$ symmetry guarantees the other three pairs of the straight nodal lines also are RNLs.
Since the RNL  constructed by the straight nodal lines  are linked by the crossed nodal lines along the two inequivalent ZR paths, we term it as crossed RNL.
Apparently, the crossed RNL  phase in AuBr is completely different from that in ABC-stacked graphdiyne, and show distinguished properties as discussed below.

\begin{figure}[b]
\includegraphics[width=0.48\textwidth]{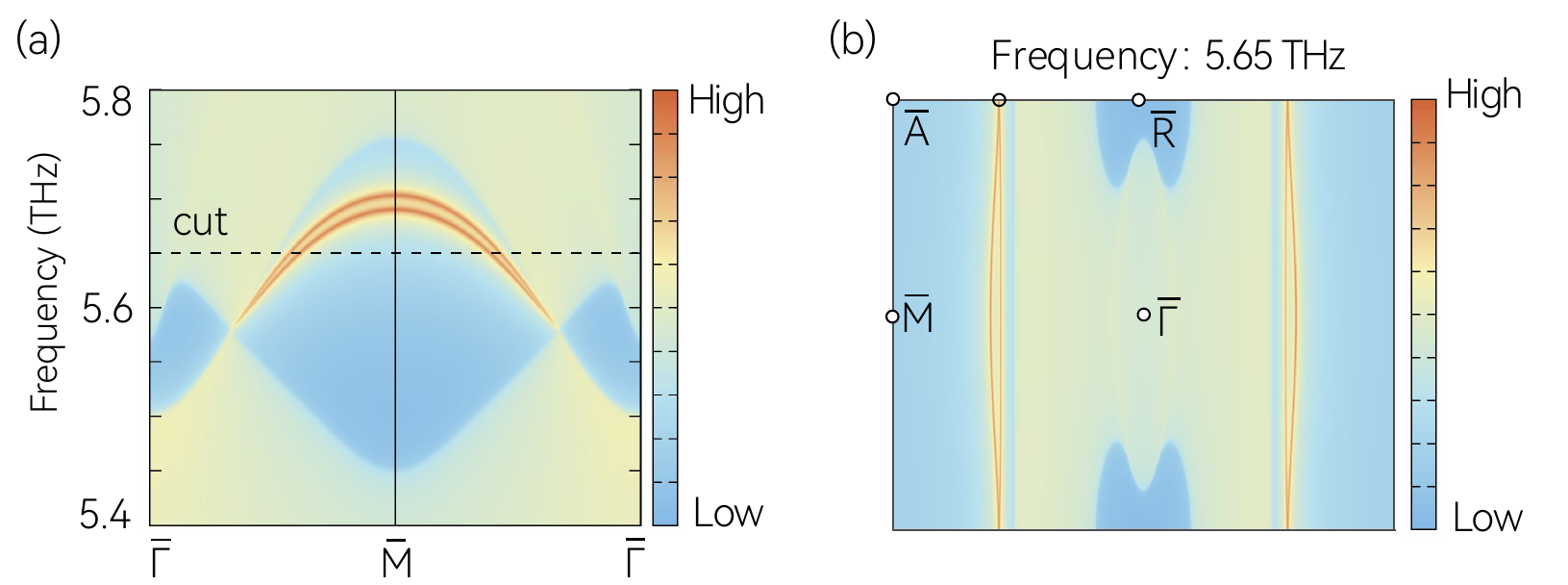}% Here is how to import EPS art
\caption{\label{surface}  (a) The surface local density of states of the phonons on the $(010)$ surface along path $\bar{\Gamma}$-$\bar{M}$-$\bar{\Gamma}$.  (b) The isofrequency arcs cut at 5.65 THz. }
\end{figure}

\section{Topological surface and hinge states}
\subsection{Topological surface  state}
For (010) and (100) surfaces which are connected by $C_{4z}$, they should have even surface states in the gap between 20$th$ and 21$th$ bands, as there are always two straight nodal lines that are projected into the same position of the corresponding surface BZ.
This is confirmed by our calculations on the surface local density of states on (010) surface, as shown in Fig. \ref{surface}, where two surface states connecting the straight nodal lines can be clearly observed.

Interestingly, the two surface states are degenerate at $\bar{R} \bar{A}$ path with $k_z = \pi$ [see  Fig. \ref{surface}(b)].
A generic point in the $\bar{R} \bar{A}$ path of the (010) surface has a combined symmetry $G_x \mathcal {T}$ with $G_x = \{M_{x}|00\frac{1}{2}\}$.
Since
\begin{equation}
\begin{aligned}
(G_{x} \mathcal {T})^2 &= \mathcal {T}^2 \{M_{x}|00\frac{1}{2}\}\{M_{x}|00\frac{1}{2}\}\\
& = \{E|001\},\\
\end{aligned}
\end{equation}
with $E$ the identity operator, one has
\begin{equation}
(G_x \mathcal {T})^2  =-1,
\end{equation}
for  $k_z = \pi$ line, i.e. $\bar{R} \bar{A}$ path.
Thus, all the bands at $\bar{R} \bar{A}$ path are at least doubly degenerate, due to the Kramers-like degeneracy.

\subsection{Topological hinge state}

\begin{figure}[t]
\includegraphics[width=0.45\textwidth]{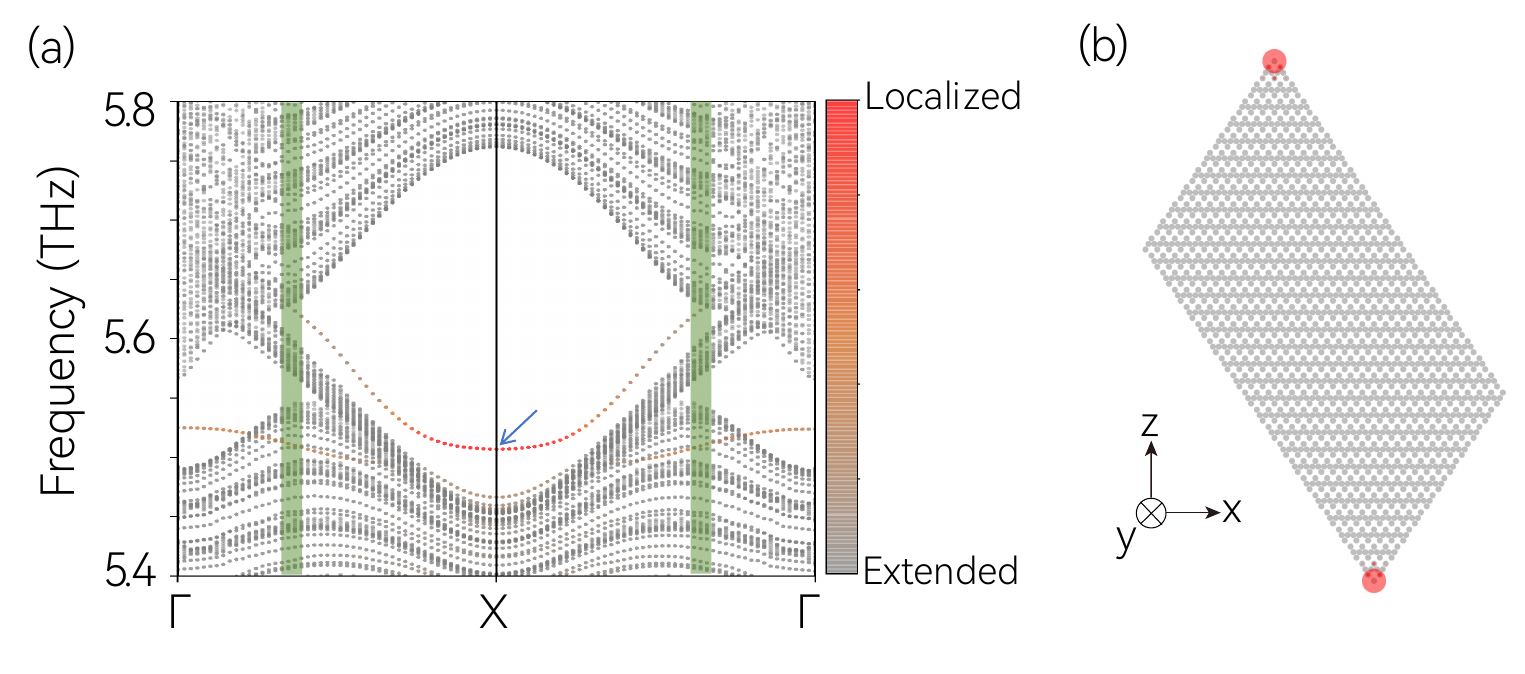}% Here is how to import EPS art
\caption{\label{hinge} (a) Spectrum for a tubelike geometry sample. The crossed RNL is located in the green shaded area.
A doubly degenerate topological hinge state connecting the  two crossed RNLs can be clearly observed.
%The color change from red to gray indicates the gradual decrease in the localization of the phonon mode.
(b) The charge density distribution of the hinge state at  $X$ point  [marked by blue arrow in (a)] in real space.
}
\end{figure}
Since the AuBr has crossed RNL in its phonon spectrum, one can expect that in addition to the 2D surface states, the AuBr also should  have topologically protected 1D hinge states.

The position  of the hinge state can be inferred from the $Z_2$ invariant $\nu_R$ for the 2D slice in bulk BZ.
As aforementioned, the   $k_y=\pi$ plane is nontrivial with $\nu_R=1$ [see Fig. \ref{wilsonloop}(b)]. Then, all the planes within  $k_y\in[-\pi,-k_c)$ and $k_y\in(k_c,\pi]$ ($k_c\simeq 0.38\pi$) would have $\nu_R=1$, as the 20$th$ and 21$th$ phonon bands are gapped in this region.
Similarly, the  planes within  $k_y\in(-k_d,k_d)$  ($k_d\simeq 0.33\pi$) would have $\nu_R=0$.
The  interval of $k_y\in(k_d,k_c)$  is the location  of the crossed RNL.
The slices with $\nu_R=1$ are 2D real Chern insulators and then must exhibit at least a pair of ${\cal{PT}}$ connected corner states.
All these corner states constitute the hinge state of the AuBr.
To explicitly demonstrate it, we calculate the spectrum of the AuBr with a tubelike geometry shown in Fig. \ref{hinge}(b), which is finite in the $x$-$z$ plane but periodic in the $y$ direction.
The result is plotted in Fig. \ref{hinge}(a), where the color denotes the degree of locality of the bands.
One observes that  a pair of localized states  exist in the interval  $k_y\in[-\pi,-k_c)$ and $k_y\in(k_c,\pi]$ and disappear when $k_y\in(-k_d,k_d)$.
By further checking the spatial distribution for these state  at X point ($k_y=\pi$), we find they are exactly corner states [see Fig. \ref{hinge}(b)].
These corner states  constitute the hinge state of the AuBr [see Fig. \ref{hinge}(a)].

In contrast to all the previous RNL states, in which the hinge state only appears in one direction, the crossed RNL in AuBr can feature hinge states in two directions.
As shown in Fig. \ref{CRNL of AuBr}(b) , the two straight nodal lines   circled in blue in $k_x$ direction also form a RNL. From $C_{4z}$ symmetry in AuBr, one easily knows that  all the planes within  $k_x\in[-\pi,-k_c)$ and $k_x\in(k_c,\pi]$ ($k_c\simeq 0.38\pi$)  also are 2D real Chern insulators with $\nu_R=1$.
Thus, for the AuBr with a similar tubelike geometry [see Fig. \ref{hinge}(b)] but being periodic in $x$ direction, it also should  exhibit a hinge state that is similar to the hinge state in Fig \ref{hinge}, and then is not explicitly shown here.

\section{Transformation of the crossed RNL under stain}
\begin{figure}[t]
\includegraphics[width=0.45\textwidth]{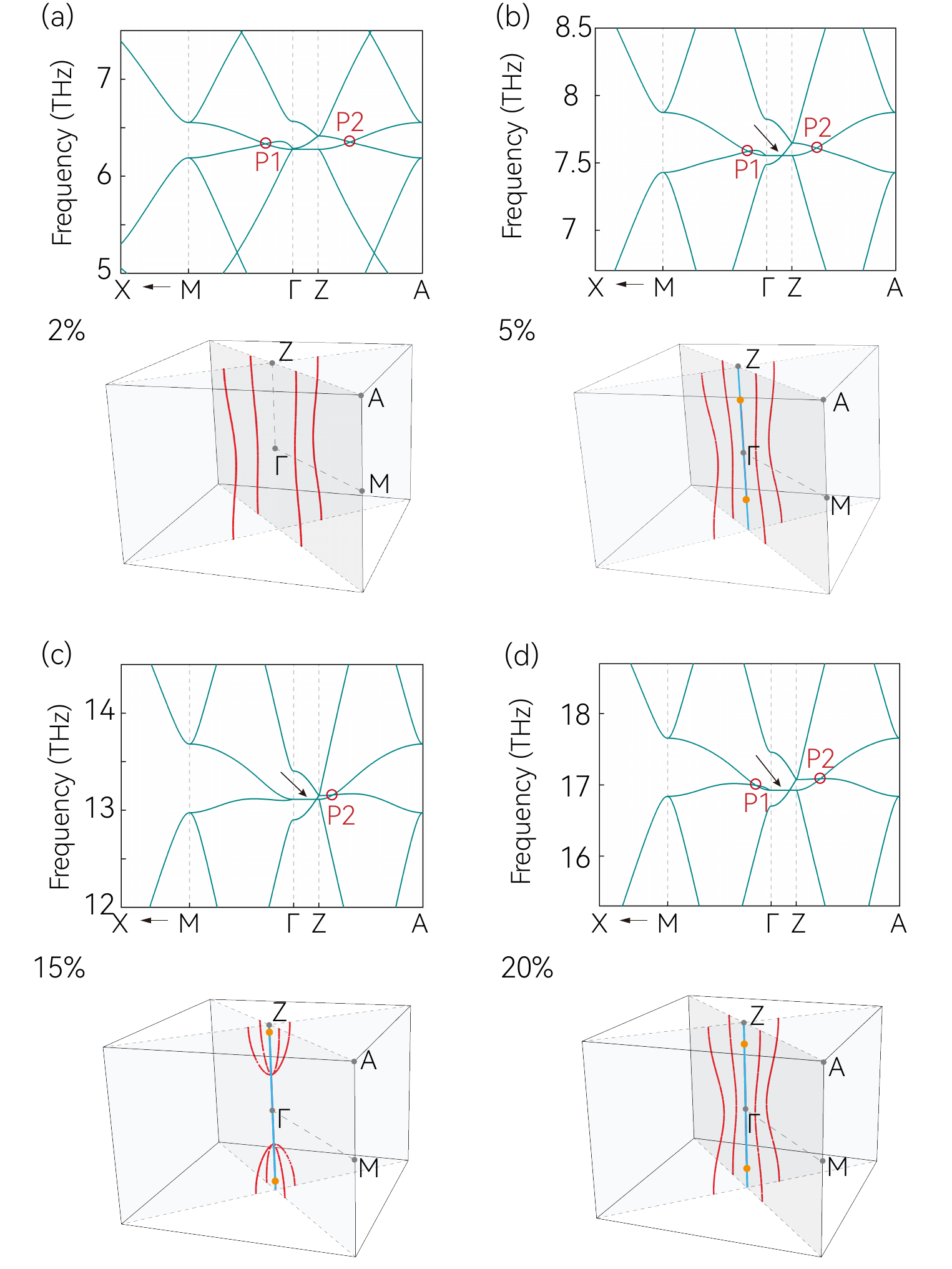}% Here is how to import EPS art
\caption{\label{strain} The phonon spectrums along path X-M-$\Gamma$-Z-A and the shape of RNLs in k-space under the (a) $2 \%$, (b) $5 \%$, (c) $15 \%$, and (d) $20 \%$ compressive stress, respectively. Here, the blue lines represent the nodal line formed by the two-degenerated bands on path $\Gamma$Z, and the orange dot represents the triple-degenerated point formed on path $\Gamma$Z. }
\end{figure}

Finally, we study the transformation of the crossed RNL in AuBr under symmetry-persevering strain.
Since the pair of the crossed RNLs are well separated in momentum space, they are robust against the small perturbations.
By applying a compressive stress, the four inequivalent $P_1$ points at $k_z=0$ plane will move toward $\Gamma$ point, making the four straight nodal line distorted [see Fig. \ref{strain}(a) and \ref{strain}(b)].
However, each adjacent two of the four nodal lines still form a crossed RNL.
When the critical stress is of $2\%$, the gap at point $\Gamma$ between the two middle bands (20$th$ and 21$th$ bands) closes [see Fig. \ref{strain}(a)], and a band inversion occurs as increasing compressive stress, resulting in a nodal line formed by 20$th$ and 21$th$ bands and a triply degenerate point \cite{lenggenhager2021triplepoint,lenggenhager2022universal,lenggenhager2022triplea} on path $\Gamma$Z [see Fig. \ref{strain}(b)].
Keep increasing the stress, the four inequivalent $P_1$ points continue to move toward $\Gamma$ point and merge into the nodal line on $\Gamma$Z path. Meanwhile, the four nodal lines become a crossed nodal loop, as shown in Fig. \ref{strain}(c). As the stress keeping increase, the crossed nodal loop shrinks but will not disappear. This is because the $P_2$ point at AZ path is a movable but unremovable point, as demonstrated by Li et. al. \cite{li2023upper}.
Thus, the crossed nodal loop will expand again when the stress continues to increase.
Remarkably, the crossed RNLs reappear around the $20\%$ compressive stress as shown in Fig. \ref{strain}(d) and Fig. \ref{rectangular}(b).

\section{Discussions}
In this work, we propose the concept of crossed RNL and predict its realization in the phonon spectrum of 3D AuBr. The configuration of the crossed RNL is completely different from the previously reported RNL. Similar to the conventional RNL, We show that the crossed RNL also exhibit two topological invariants: $\pi$ Berry phase and real Chern number, then possessing  both topological surface states and hinge states. However, the topological boundary state in 3D AuBr is distinguished from that of the conventional RNL.
Moreover, the crossed RNL in AuBr is robust against the symmetry-persevering perturbations and show unique transformation under uniform strain.

Besides phonon spectrum, the crossed RNL also can appear in the electronics bands of the materials with negligible spin-orbit coupling.
Moreover, the artificial systems such as acoustic and photonic crystals may be promising platforms for realizing the crossed RNL, due to the highly flexible in tuning hopping parameters \cite{xue2019realization,zhang2019dimensional,luo2021observation,wei20213d,wei2021higherorder,pu2023acoustic,pan2023real}.

\acknowledgements
The authors thank Si Li and J. Xun for helpful discussions.
This work is supported by the National Key R$\&$D Program of China (Grant No. 2020YFA0308800), the Science Fund for Creative Research Groups of NSFC (Grant No. 12321004), and the National Science Foundation of China (Grant Nos. 12234003 and 12004035). Yilin Han and Yichen Liu contributed equally to this work.

\appendix

\section{Calculation Method}
The first-principles calculation of 3D AuBr were conducted via the Vienna ab initio simulation package (VASP) \cite{kresse1996efficient,kresse1996efficiency}. The projector augmented wave (PAW) pseudopotentials \cite{blochl1994projector,kresse1999ultrasoft} were adopted in the calculation and generalized gradient approximation (GGA) \cite{perdew1996generalized} of the Perdew-Burke-Ernzerhof (PBE) functional \cite{perdew1998perdewa} was selected as the exchange correlation potential. The 2D Brillouin zone (BZ) was sampled using Monkhorst-Pack k-mesh \cite{monkhorst1976special} with a size of 11$\times$11$\times$4. The energy cutoff was set to 600 eV and the energy convergence criteria was $10^{-8}$ eV. The phonon band structures of AuBr was calculated by the PHONOPY package \cite{togo2015firsta}. Furthermore, based on the first-principles calculation, the Wannier tight-binding (TB) model for the low-energy bands of 3D AuBr was constructed by Wannier90 \cite{mostofi2008wannier90,mostofi2014updated} for the subsequent higher-order band topology calculation. The surface states and the scanning of the crossed RNL are both displayed via Wanniertools \cite{wu2018wanniertools}.

The electronic band structure of AuBr with and without spin-orbit coupling (SOC) is calculated via NANODCAL package \cite{taylor2001initio}.

\section{The Wilson loop of the Crossed RNL with and without compressive strain}
\begin{figure}[b]
\includegraphics[width=0.48\textwidth]{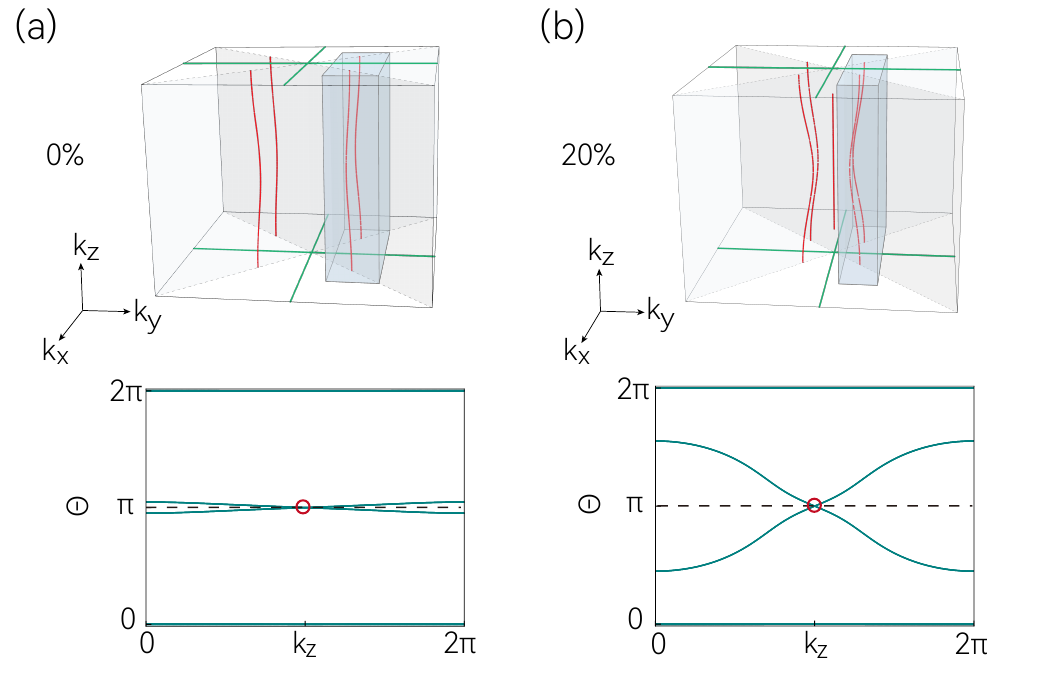}% Here is how to import EPS art
\caption{\label{rectangular} Wilson loop of the 17th to 20th bands calculated on a 2D torus (marked as blue), which includes a pair of straight nodal line. (a) shows the Wilson loop for the 3D AuBr  without strain and  (b) shows that with $20\%$ compressive stress. }
\end{figure}
As discussed in the main text, the switch in topology with $k_y$ from $\nu_R = 0$ to $\nu_R = 1$ dictates the existence of crossed RNL between the two planes.
It also follows that each RNL carries this nontrivial 2D topological charge $\nu_R = 1$ defined on a 2D torus  surrounding the RNL, as shown in Fig. \ref{rectangular}.
Figures \ref{rectangular}(a) and  \ref{rectangular}(b) show the calculated the Wilson  loop on the 2D torus of the 3D AuBr  without strain and  with $20\%$ compression stress, respectively.
The odd (one) crossing point at $\theta =\pi$ in  Fig. \ref{rectangular} demonstrate  that the crossed RNL exists in both cases.

\section{Electronic band structure of AuBr}

\begin{figure}[h]
\includegraphics[width=0.45\textwidth]{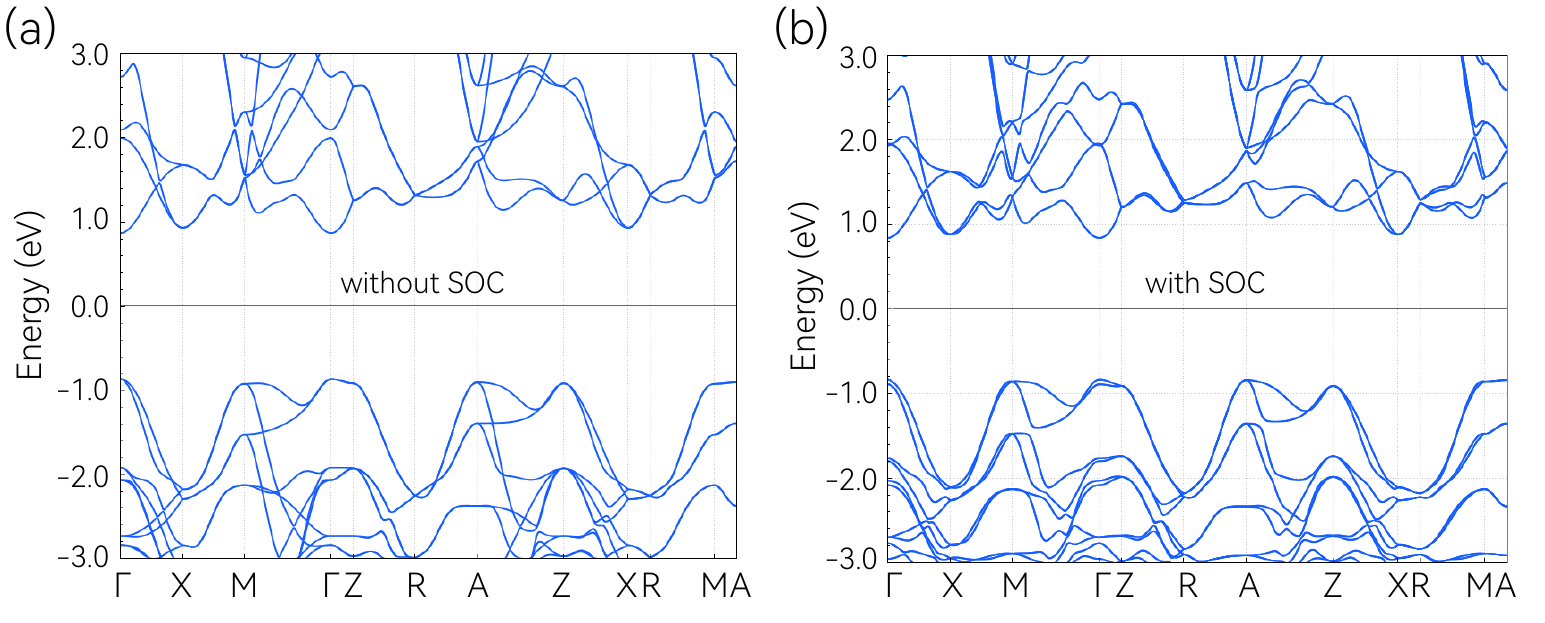}% Here is how to import EPS art
\caption{\label{band} Band structure of the 3D AuBr  (a) without SOC and (b) with SOC. }
\end{figure}

In addition to the phonon spectrum, the crossed RNL could theoretically also appear in the electronic bands of the AuBr without SOC, as the two spectrum respect the  symmetry of  ${\cal{PT}}^2=1$.
Here, we calculate the electron bands of 3D AuBr without and with SOC [see Fig. \ref{band}(a) and (b)]. However, the bands are relatively messy and no obvious RNL band structure is found near the Fermi level.

\bibliography{mybib}% Produces the bibliography via BibTeX.
\end{document}